\documentclass[journal]{IEEEtran}
\usepackage{amsmath,amssymb,amsfonts}
\usepackage{graphicx}
\usepackage{algorithm,algorithmic}
\usepackage{hyperref}
\hypersetup{hidelinks=true}
\usepackage{textcomp}
\usepackage{booktabs} 
\usepackage{xcolor}
\usepackage{cleveref}
\usepackage{bm}         

\hypersetup{
    colorlinks=true,
    urlcolor=magenta  
}
\definecolor{fan}{RGB}{255, 100, 0}

\def\BibTeX{{\rm B\kern-.05em{\sc i\kern-.025em b}\kern-.08em
    T\kern-.1667em\lower.7ex\hbox{E}\kern-.125emX}}
\begin{document}
\title{MTMed3D: A Multi-Task Transformer-Based Model for 3D Medical Imaging}
\author{Fan Li, Arun Iyengar, Lanyu Xu
\thanks{This work is supported by the National Science Foundation (NSF) under Grant No. 2245729. 
}
\thanks{Fan Li is with the Department of Computer Science and Engineering, Oakland University, Rochester Hills, MI USA
(e-mail: fanli@oakland.edu). }
\thanks{Arun Iyengar is with Intelligent Data Management and Analytics, LLC, Yorktown Heights, NY USA
(e-mail: aki@akiyengar.com). }
\thanks{Lanyu Xu is with the Department of Computer Science and Engineering, Oakland University, Rochester Hills, MI USA
(e-mail: lxu@oakland.edu). }
}
\maketitle

\begin{abstract}
In the field of medical imaging, AI-assisted techniques such as object detection, segmentation, and classification are widely employed to alleviate the workload of physicians and doctors. However, single-task models are predominantly used, overlooking the shared information across tasks. This oversight leads to inefficiencies in real-life applications. 
In this work, we propose MTMed3D, a novel end-to-end Multi-task Transformer-based model to address the limitations of single-task models by jointly performing 3D detection, segmentation, and classification in medical imaging. Our model uses a Transformer as the shared encoder to generate multi-scale features, followed by CNN-based task-specific decoders. The proposed framework was evaluated on the BraTS 2018 and 2019 datasets, achieving promising results across all three tasks, especially in detection, where our method achieves better results than prior works. Additionally, we compare our multi-task model with equivalent single-task variants trained separately. Our multi-task model significantly reduces computational costs and achieves faster inference speed while maintaining comparable performance to the single-task models, highlighting its efficiency advantage. To the best of our knowledge, this is the first work to leverage Transformers for multi-task learning that simultaneously covers detection, segmentation, and classification tasks in 3D medical imaging, presenting its potential to enhance diagnostic processes. The code is available at \url{https://github.com/fanlimua/MTMed3D.git}.
\end{abstract}


\section{Introduction}
\label{sec:intro}
In the real world, medical diagnoses are often complicated and require assessing tumor morphology.
For example, developing a treatment plan requires understanding the location, distinguishing between healthy and diseased tissue, and classifying disease type and severity. These steps rely heavily on medical imaging, making tasks like object detection, segmentation, and classification crucial for diagnoses. With advancements in deep learning, its application in the medical field has significantly improved diagnostic accuracy and efficiency.
However, most existing models in medical image analysis are single-task models designed to perform only one task at a time. While single-task models are effective for addressing individual tasks, they face several limitations when compared to multi-task models. 
In particular, they may struggle to generalize when trained on small datasets, which is common in medical imaging where labeled data is often limited.
Furthermore, the single-task design isolates them from one field, preventing them from leveraging knowledge from related tasks \cite{vandenhende2021multi}. In contrast, multi-task models share representations across tasks, enabling them to learn complementary features and utilize information that may not be directly available in the primary task, thereby addressing the data scarcity challenge more effectively. Additionally, single-task models require redundant computations and increased resource consumption for each task, making them inefficient compared to multi-task models. The multi-task models share features and parameters across tasks, allowing them to perform multiple tasks simultaneously. This reduces computational costs, making the model more efficient and practical for medical image analysis, where computational resources are often limited.

In recent years, Transformer-based models have shown promising results for medical image analysis \cite{chen2021transunet, hatamizadeh2022unetr, cao2022swin, wenxuan2021transbts, hatamizadeh2021swin}. Some of these models leverage self-attention and positional encodings, enabling them to effectively capture long-range dependencies. However, traditional transformers like Vision Transformer (ViT) \cite{dosovitskiy2020image} suffer from high computational complexity due to the quadratic scaling of global self-attention, making them inefficient. To address this, Swin Transformers \cite{liu2021swin} apply self-attention within smaller, localized windows and then use the shift window-based multi-head self-attention module to establish connections between neighboring non-overlapping windows. This enables Swin Transformers to capture local and global context efficiently without the heavy computational cost. Building on Swin Transformer advancements, we propose an end-to-end Transformer-based multi-task model. While brain tumors serve as the test case for evaluating MTMed3D's performance, its architecture can be adapted to other medical imaging tasks.
\begin{figure*}[!ht]
\centering
\includegraphics[width=0.95\textwidth]{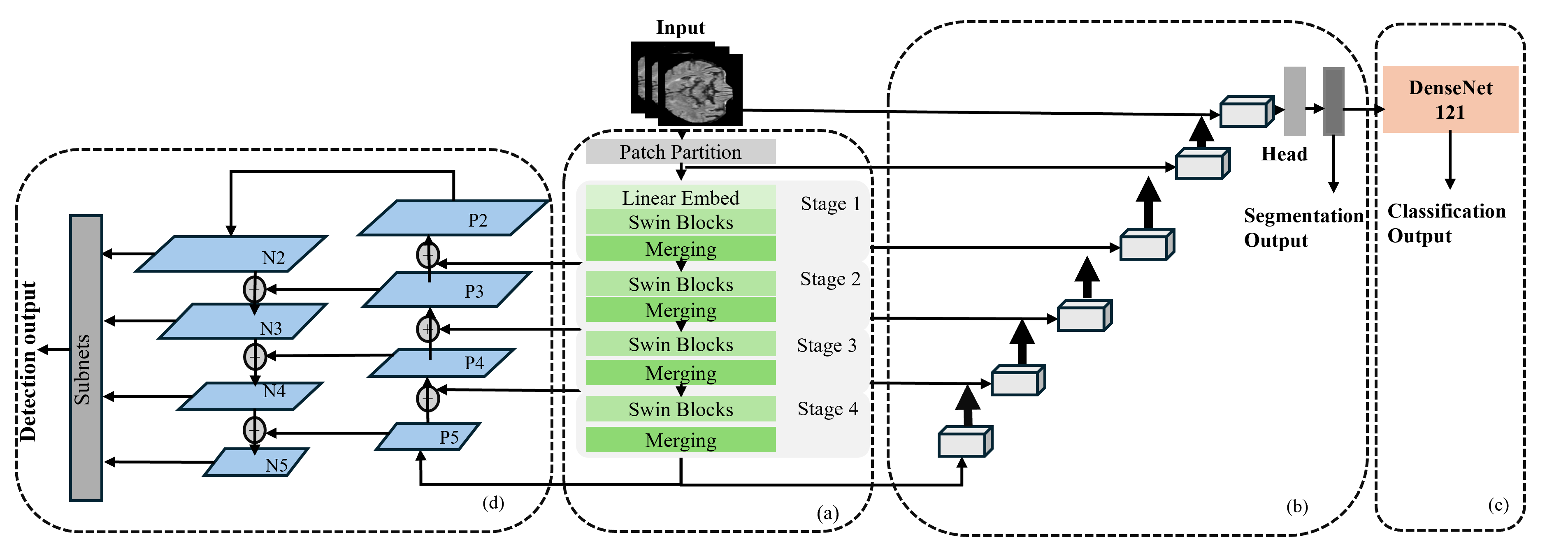}
\caption{Overview of MTMed3D: (a) Swin Transformer-based encoder, (b) Segmentation Decoder, (c) Classification Decoder, (d) Detection Decoder. The input to our model is 3D multi-modal MRI images with 4 channels. The data first passes through a shared Swin Transformer-based encoder and then flows through three task decoders to generate results for all three tasks.}
\label{fig1_architecture}
\end{figure*}
In summary, our main contributions are as follows:
\begin{itemize}
    \item We propose MTMed3D, the first end-to-end Multi-task Transformer-based model for 3D medical image segmentation, detection, and classification.
    \item We evaluated MTMed3D on the Multimodal Brain Tumor Segmentation Challenge (BraTS) \cite{menze2014multimodal} 2018 and 2019, showing its promising performance across all three tasks, with detection performance surpassing that of existing methods. We have open-sourced MTMed3D at \href{https://github.com/fanlimua/MTMed3D.git}{https://github.com/fanlimua/MTMed3D.git}. 
    \item We show that MTMed3D outperforms single-task models in efficiency with lower computation and latency. Additionally, we conduct a comprehensive ablation study, validating the effectiveness of our model design.
\end{itemize}

\section{RELATED WORK}
\subsection{Multi-task Learning}
Multi-task learning (MTL) is a paradigm in machine learning that aims to jointly learn multiple related tasks, leveraging the knowledge from one task to benefit others. This approach hopes to improve the generalization performance of all tasks involved \cite{zhang2021survey}. Currently, multi-task learning is applied in many fields of machine learning, such as autonomous driving. YOLOP \cite{wu2022yolop} proposed an efficient multi-task model based on CNN, which can simultaneously handle traffic object detection, drivable area segmentation, and lane detection. Multi-task learning has also been applied to the field of medical images. Park et al. \cite{park2021federated} proposed a multi-task model based on a ViT, which can perform three tasks of lung image classification, target detection, and segmentation at the same time. Their approach differs from ours in several key aspects. They utilize federated learning and split learning, training the shared transformer body centrally and task-specific heads and tails on separate machines with different datasets. In contrast, our model is trained end-to-end on a single machine using the same image for all tasks. Additionally, they employ a ViT as the shared encoder, which has a significantly larger parameter count of 66.367 million compared to the 8.063 million parameters in our Swin Transformer encoder. Amyar et al. \cite{amyar2020multi} present an automatic classification segmentation tool for helping to diagnose COVID-19 pneumonia using chest CT imaging. Yan et al. \cite{yan2019mulan} introduced a multi-task universal lesion analysis network designed for the joint detection, tagging, and segmentation of lesions across various body parts. This network is built upon an improved Mask R-CNN framework featuring three head branches and a 3D feature fusion strategy.

\subsection{Transformer Models and Their Applications}
Transformer was proposed by \cite{vaswani2017attention} and was initially used for machine translation and natural language processing (NLP) tasks. Its success has been widely attributed to the self-attention mechanism, which effectively models long-range dependencies \cite{bahdanau2014neural, cheng2016long}. Given this capability, Transformer has been increasingly adopted in computer vision tasks. ViT \cite{dosovitskiy2020image} processes images as tokenized patches for classification but suffers from high computational complexity at high resolutions. To address this, Swin Transformer \cite{liu2021swin} introduces window attention, applying local attention within windows. This method effectively reduces complexity. Transformer-based models have gained increasing attention in medical image analysis \cite{shamshad2023transformers, liu2023recent}. For segmentation, TransUNet \cite{chen2021transunet} was the first Transformer-based framework for medical image segmentation, integrating a Transformer layer into the bottleneck of a U-Net for multi-organ segmentation. Swin UNETR \cite{hatamizadeh2021swin} leverages a Swin Transformer encoder to extract rich semantic features and a CNN decoder to reconstruct structured outputs. Transformer-based architectures have also been explored for classification tasks. Lu et al. \cite{lu2021smile} proposed a two-stage framework that first employs contrastive pre-training to extract meaningful representations and then aggregates features using a Transformer-based sparse attention module for label prediction. Zhang et al. \cite{zhang2021transformer} introduced a Transformer-based two-stage framework for COVID-19 diagnosis in 3D CT scans. Their approach involves using a UNet for lung segmentation, followed by Swin Transformer feature extraction from each CT slice, which is then aggregated into a 3D volume-level representation for classification. 
\begin{figure*}[tbp]
\centering
\includegraphics[width=0.75\textwidth]{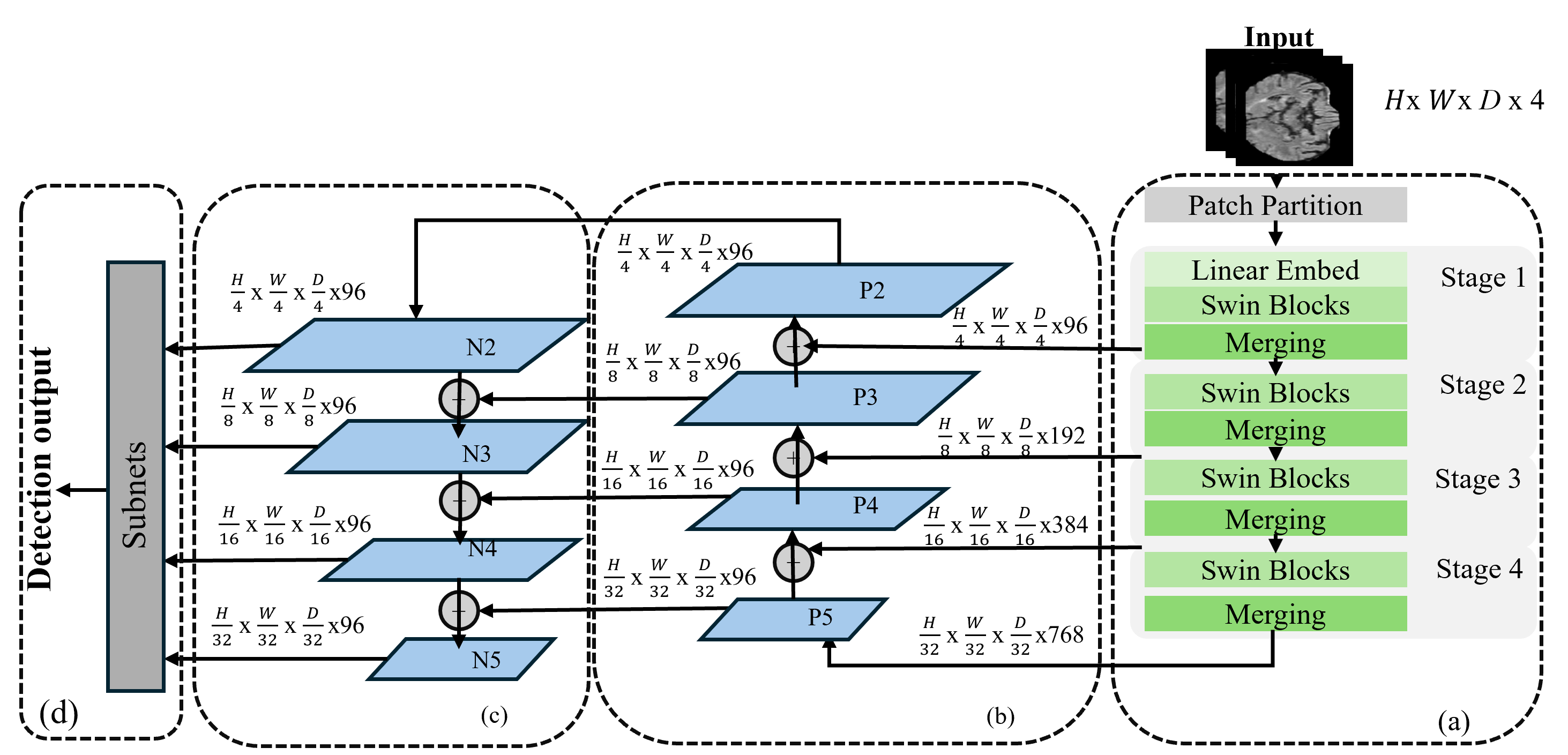}
\caption{Detection Network: (a) Swin Transformer-based encoder, (b) FPN, (b) + (c) PANet, (d) Subnets for final box prediction are composed of Convs, GroupNorm, and ReLU.}
\label{fig6_detection}
\end{figure*}
For object detection tasks, Shen et al. \cite{shen2021cotr}, inspired by the Detection Transformer (DETR), proposed a Convolutional Transformer (COTR) network for end-to-end polyp detection. COTR consists of a CNN for feature extraction, Transformer encoder layers interleaved with convolutional layers for feature encoding and recalibration, Transformer decoder layers for object querying, and a feed-forward network for detection prediction. This architecture effectively leverages the strengths of both CNNs and Transformers, enabling COTR to outperform DETR on two different datasets. Building on previous research, we explore the potential of the Transformer as a shared backbone for multi-task medical imaging.
\section{MTMed3D Model Design}
\subsection{Design Overview}
We propose MTMed3D, a Multi-task Transformer-based model for 3D medical image analysis that consists of a shared encoder and three decoders. An overview of the proposed MTMed3D is presented in \Cref{fig1_architecture}. 
In designing this multi-task model, we solved the following challenges. {\bf \textit{Challenge 1:}} Selection of appropriate model architecture. {\bf Solution:} We adopted hard parameter sharing \cite{meyerson2017beyond}, enabling tasks to use a shared encoder. {\bf \textit{Challenge 2:}} Task balancing among multiple tasks. {\bf Solution:} We used Gradient Normalization (GradNorm) \cite{chen2018gradnorm} to adjust loss weights across tasks. {\bf \textit{Challenge 3:}} The absence of detection labels. {\bf Solution:} We generated detection labels from existing segmentation annotations.
\subsection{Encoder}
Our multi-task model uses a shared Transformer encoder implemented with Swin Transformer blocks, which consist of four hierarchical stages. Each stage consists of Swin blocks followed by a patch merging layer. Given a 3D image, we first apply a patch partition layer to create flattened 3D patches, followed by a linear embedding layer to project the patches into a latent space. The patch embeddings then pass through four Swin Transformer blocks and patch merging layers to generate hierarchical feature representations. 
In the Swin Transformer block, the window-based multi-head self-attention (W-MSA) module computes self-attention within local windows, which helps to reduce computational complexity. However, W-MSA lacks connections across windows, so the shift window-based multi-head self-attention (SW-MSA) module is applied to establish connections between neighboring non-overlapping windows in the previous layer. After SW-MSA, mutual communication between windows can be achieved. 
So Swin Transformer block \cite{liu2021swin} can be formulated as:
\begin{align}
    z_{l}^{\prime} &= W-MSA(LN(z_{l-1})) + z_{l-1}, \\
    z_l &= MLP(LN(z_{l}^{\prime})) + z_{l}^{\prime}, \\
    z_{l+1}^{\prime} &= SW-MSA(LN(z_l)) + z_l, \\
    z_{l+1} &= MLP(LN(z_{l+1}^{\prime})) + z_{l+1}^{\prime}.
\end{align}
\(z_l^{\prime}\) and \(z_{l+1}^{\prime}\) denote the outputs of W-MSA and SW-MSA; MLP and LN denote Multi-Layer Perceptron and Layer Normalization, respectively. Similar to prior works \cite{vaswani2017attention}, self-attention is computed using queries, keys, and values. After each Swin Transformer block, a patch merging layer reduces the resolution of feature maps, enabling the encoder to generate multi-scale features. This shared encoder is used across tasks to address Challenge 1, where tasks share parameters and representations to reduce redundant computation.

\subsection{Decoders}
In this section, we introduce three decoders, each designed to handle a different task.
\subsubsection{Detection Decoder}
\begin{figure*}[tbp]
\centering
\includegraphics[width=0.9\textwidth]{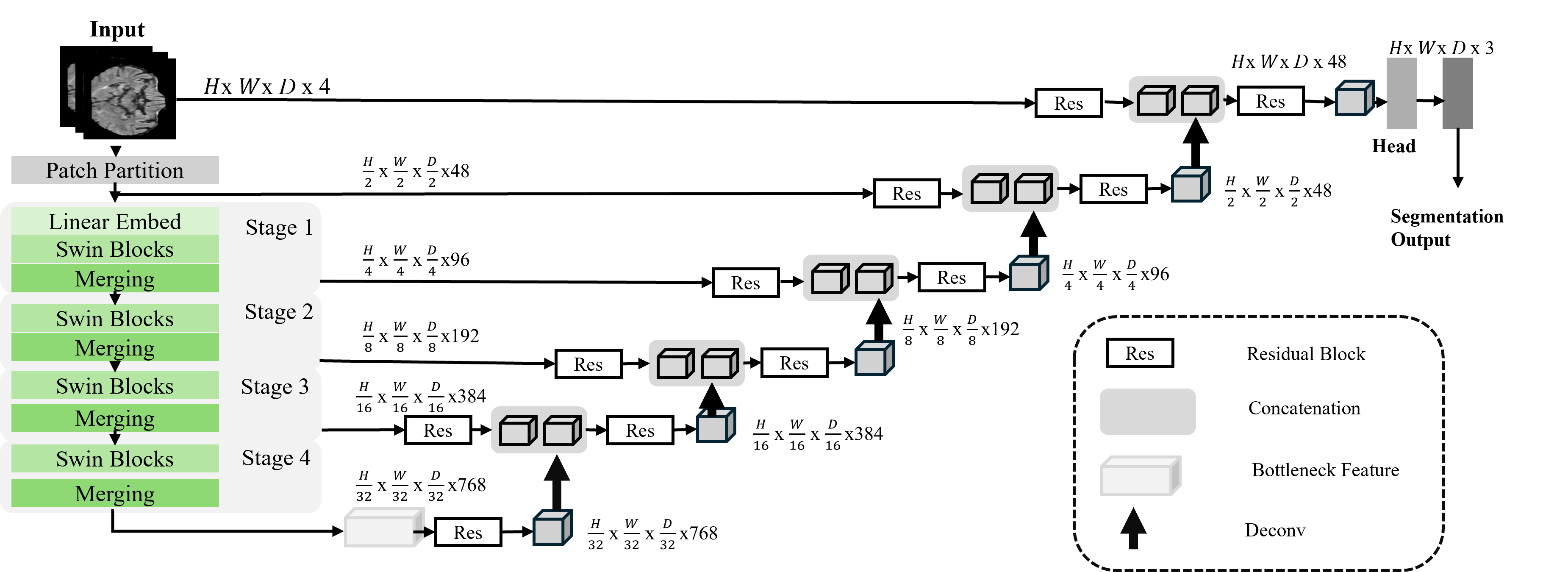}
\caption{Segmentation Network: The Swin Transformer-based encoder and CNN-based decoder form a U-shape segmentation network.}
\label{fig4_segmentation}
\end{figure*}
For the detection task, we utilize a modified RetinaNet \cite{lin2017focal} framework. In our modified RetinaNet, we adopt the Swin Transformer encoder as the backbone to extract features and replace the Feature Pyramid Networks (FPN) \cite{lin2017feature} architecture with Path Aggregation Network (PANet) \cite{liu2018path}, as shown in \Cref{fig6_detection}. This approach has not been proposed in previous works. The Swin Transformer encoder extracts multi-scale feature maps at different stages $i$ (i=2,3,4) and the bottleneck (i=5), which are reshaped to a uniform channel size using convolutional blocks. These reshaped features are progressively upsampled and fused with the previous layer’s reshaped features through skip connections. Applying convolution filters on these features to generate the feature maps $P_{i}$ (i=2,3,4,5) for PANet. PANet extends FPN by introducing an additional pathway as (c) in \Cref{fig6_detection} to enhance low-level feature propagation. In this path, Each feature map $N_{i}$ first goes through a convolutional layer to reduce the spatial size and concatenate with feature map $P_{i+1}$ via a skip connection. The fused feature map is processed by a convolutional layer to generate $N_{i+1}$ for the following sub-networks. The box regression subnet uses convolutional layers, GroupNorm, and ReLU to predict object coordinates.

\subsubsection{Segmentation Decoder}
The U-Net architecture has shown promising results in medical image segmentation, leading many recent studies \cite{li2018h, zhou2018unet++, zhou2019unet++, jha2020doubleu, shang2024lk} to design their architectures in a U-Net shape. So in our segmentation decoder, we follow the Swin UNETR \cite{hatamizadeh2021swin} network to achieve segmentation. Swin UNETR extracts the feature representation using the Swin Transformer encoder and feeds them to the decoder via skip connection at each stage, as shown in \Cref{fig4_segmentation}. Features from each encoder stage i (i=0,1,2,3,4,5) are processed through residual blocks. The decoder progressively increases resolution by deconvolution layers, concatenating features from previous stages to refine segmentation. A final convolutional layer with a sigmoid activation generates the segmentation output. This U-shaped architecture effectively integrates multi-scale features, enhancing segmentation detail and performance.

\subsubsection{Classification Decoder}
Tripathi et al. \cite{tripathi2022attention} used segmentation outputs as inputs for classification, leveraging detailed tumor information to improve performance. Following this approach, we place our classification branch after segmentation, using DenseNet-121 \cite{huang2017densely} for tumor grading.
DenseNet improves information flow by connecting each layer directly to all subsequent layers.
Hence, the $l$\-th layer receives the feature maps of all preceding layers, and the formula is computed as follows: 
\begin{equation}
x_{l} = H_{l}([x_0, x_1, \ldots, x_{(l-1)}])   
\end{equation}
where \([x_0, x_1, \ldots, x_{(l-1)}]\) represents the concatenation of the feature maps produced in layers \( 0, 1, 2, \ldots, l-1 \), and $H_{l}$ can be a composite function of operations, $x_{l}$ denotes the output of the $l$\-th layer. DenseNet can alleviate gradient vanishing and reduce model size.
\begin{table*}[htbp]
  \centering
  \caption{Segmentation, Classification, and Detection Results on Five-Fold Cross-Validation.}
  \resizebox{\textwidth}{!}{ 
    \begin{tabular}{cccccccccccccc}
      \toprule
      \textbf{Task} & \multicolumn{6}{c}{\textbf{Segmentation}} & \multicolumn{3}{c}{\textbf{Classification}} & \multicolumn{4}{c}{\textbf{Detection}} \\
      \cmidrule(r){2-7} \cmidrule(r){8-10} \cmidrule(r){11-14}
      \textbf{Metric} & \multicolumn{3}{c}{\textbf{Dice}} & \multicolumn{3}{c}{\textbf{HD}} & \textbf{Acc} & \textbf{Sen} & \textbf{Spe} & \textbf{mAP@} & \textbf{mAP@} & \textbf{mAR@} & \textbf{mAR@}\\
      \cmidrule(r){2-4} \cmidrule(r){5-7}
      \textbf{Region} & \textbf{WT} & \textbf{TC} & \textbf{ET} & \textbf{WT} & \textbf{TC} & \textbf{ET} & & & & 0.1:0.5 & 0.5 & 0.1:0.5 & 0.5 \\
      \midrule
      {BraST 2018 (Best)} & 0.9082 & 0.8183 & 0.8038 & 5.4346 & 6.9773 & 5.6794 & 0.9298 & 0.9286 & 0.9333 & 0.9711 & 0.8650 & 0.9844 & 0.9298\\
      {BraST 2018 (Avg.)} & 0.8793 & 0.8019 & 0.7755 & 9.7864 & 8.3361 & 6.7452 & 0.9193 & 0.9761 & 0.7634 & 0.9082 & 0.7628 & 0.9357 & 0.8421\\
      \midrule
      {BraST 2019 (Best)} & 0.8814 & 0.8084 & 0.7879 & 13.1085 & 9.9521 & 7.7802 & 0.9701 & 0.9804 & 0.9375 & 0.8820 & 0.7368 & 0.9187 & 0.8209\\
      {BraST 2019 (Avg.)} & 0.8875 & 0.8311 & 0.8020 & 9.2851 & 7.5083 & 5.5534 & 0.8955 & 0.9769 & 0.6153 & 0.8793 & 0.7250 & 0.9257 & 0.8269\\
      \bottomrule
    \end{tabular}
    }
  \label{tab:fivefold_reults}
\end{table*}
\begin{table}[htbp]
  \centering
  \caption{Detection performance of different methods.}
  \resizebox{\linewidth}{!}{
  \begin{tabular}{lcccc}
    \toprule
    \textbf{Methods} & \textbf{mAP@0.1:0.5} & \textbf{mAP@0.5} & \textbf{mAR@0.1:0.5} & \textbf{mAR@0.5}\\
    \midrule
    RetinaNet \cite{lin2017focal} & 0.8664 & 0.6768 & 0.8850 & 0.7193 \\
    nnDetection \cite{sobek2024medyolo}  & -- & 0.8360 & -- & -- \\
    MedYOLO \cite{sobek2024medyolo} & -- & 0.8610 & -- & -- \\
    \midrule
    MTMed3D (Best) & \textbf{0.9711} & \textbf{0.8650} & \textbf{0.9844} & \textbf{0.9298}\\
    \bottomrule
  \end{tabular}
  }
  \label{tab:det_performance}
\end{table}
\subsection{Optimization}
Our network uses three decoders, with the total loss formulated as a weighted sum of their respective loss components:
\begin{equation}
{L}_{\text{total}} = w_1 {L}_{\text{seg}} + w_2 {L}_{\text{cls}} + w_3 {L}_{\text{det}}
\end{equation}
where $Lseg$ is DiceLoss, $Lcls$ is FocalLoss, and $Ldet$ is SmoothL1Loss, $w_1$, $w_2$, and $w_3$ are the weights correspond to the three tasks. 
A key challenge in multi-task learning is balancing task performance, as one task may dominate training due to larger gradients. To address the imbalance described in Challenge 2, we employ GradNorm \cite{chen2018gradnorm}, which directly adjusts gradient magnitudes by optimizing multi-task loss. A multi-task loss function can be expressed as follows: 
\begin{equation}
L(t) = \sum w_i(t) L_i(t)
\end{equation}
where the sum runs over all $T$ tasks at training step $t$. This method first normalizes gradient norms across tasks by defining the $L2$ norm of the gradient for the weighted single-task loss $w_i(t)\times L_i(t)$ with respect to the chosen weights $w$ as $G_w^{(i)}(t)$. The mean task gradient $\bar{G_w}$ averaged across all task gradients $G_w^{(i)}$ at training step t is considered as a common basis from which the relative gradient sizes across tasks can be measured:  
\begin{equation}
\overline G_w(t) = \mathbb{E}_{\text{task}}\left[G_w^{(i)}(t)\right]
\end{equation}
To identify which tasks are lagging in the current training process, the inverse training rate $L_i$ of task $i$ at training step $t$ is defined as follows: 
\begin{equation}
\widetilde{L_i}(t) =  {L_i(t)}/{L_i(0)}
\end{equation}
To compare the training progress of different tasks, the relative inverse training rate of task $i$ at training step $t$ is defined as follows: 
\begin{equation}
r_i(t) = {\widetilde{L_i}(t)}/{\mathbb{E}_{\text{task}}\left[\widetilde{L_i}(t)\right]}
\end{equation}
When a task's loss decreases slowly, its inverse training rate $\widetilde{L_i}(t)$ will be high. This results in an increase in the relative inverse training rate $r_i(t)$. Therefore, the desired gradient norm for each task $i$ is simply:
\begin{equation}
G_w^{(i)}(t) \rightarrow \overline G_w(t) \cdot r_i(t)
\label{target}
\end{equation}
\Cref{target} gives a target for each task $i$'s gradient norms, and GradNorm updates loss weights to move gradient norms towards this target for each task. GradNorm is implemented as an $L_1$ loss function $L_{grad}$ between the actual and target gradient norms at each timestep for task $i$:
\begin{equation}
L_{\text{grad}}(i) = \left| G_w^{(i)}(t) - \overline G_w(t) \cdot r_i(t) \right|
\end{equation}
$L_{grad}$ is then differentiated only with respect to the $w_i$, as the $w_i$ directly control gradient magnitudes per task. The computed gradients are then applied via optimizer to update $w_i$.
In practice, these weights, $w_1$, $w_2$, and $w_3$, are updated iteratively during training via backpropagation. 

\section{EXPERIMENT}
\subsection{Setting}
\subsubsection{Dataset}
We evaluated our proposed model on BraTS 2018 and BraTS 2019 \cite{menze2014multimodal}, which offers different image modalities, including T1, T2, T1ce, and FLAIR. We combined four MRI modalities into a single four-channel input. The BraTS datasets provide segmentation labels for necrosis (label 1), edema (label 2), and enhancing tumor (label 4), which are grouped into three sub-regions: Whole Tumor (WT, labels 1, 2, 4), Tumor Core (TC, labels 1, 4), and Enhancing Tumor (ET, label 4). The BraTS 2018 dataset includes 210 High-Grade Glioma (HGG) and 75 Low-Grade Glioma (LGG) cases, while BraTS 2019 contains 259 HGG and 76 LGG cases. To address Challenge 3, as BraTS lacks official object detection labels, we generate bounding boxes by extracting the minimum and maximum coordinates from the segmentation annotations.

\subsubsection{Data Augmentation}
The input image size is $4 \times 240 \times 240 \times 155$, We arranged the channels as the first dimension and used the RAS (Right, Anterior, Superior) orientation. We normalized all input images and performed a center crop from the 3D image volumes to $4 \times 96 \times 96 \times 96$ to reduce the computational load. We applied a random axis mirror flip with a probability of 0.5 for all 3 axes and random rotation with a probability of 0.75. Additionally, we applied data augmentation transformers for a random per-channel intensity shift in the range (-0.1, 0.1) and random scale of intensity in the range (-0.1, 0.1) to the input channel.

\subsubsection{Implementation Details}
We conducted experiments on a machine equipped with a 13th Gen Intel Core i7-13700F CPU and an NVIDIA GeForce RTX 4070 Ti GPU. For the segmentation task, we set the learning rate to 1e-4, while for the detection and classification tasks, we set it to 1e-5. During training, we utilized a cosine annealing learning rate scheduler and the AdamW optimizer. The batch size was set to 1 to prevent excessive memory usage. Our model was trained using a five-fold cross-validation scheme with a 4:1 ratio.

\subsubsection{Evaluation Metrics}
We used task-specific metrics to evaluate our model's performance. Segmentation performance is evaluated using the Dice Coefficient (Dice) and Hausdorff Distance (HD). We used mean Average Precision (mAP) and mean Average Recall (mAR) for detection. For classification, we utilized Accuracy (Acc), Sensitivity (Sen), and Specificity (Spe). Additionally, we compared the efficiency of the multi-task model with three single-task models using metrics such as Multiply-Accumulate Operations (MACs), Floating Point Operations (FLOPs), Parameters (Params), model size, and latency.

\subsection{Results}
\subsubsection{Detection Results}
\begin{figure*}[tbp]
\centering
\includegraphics[width=0.85\textwidth]{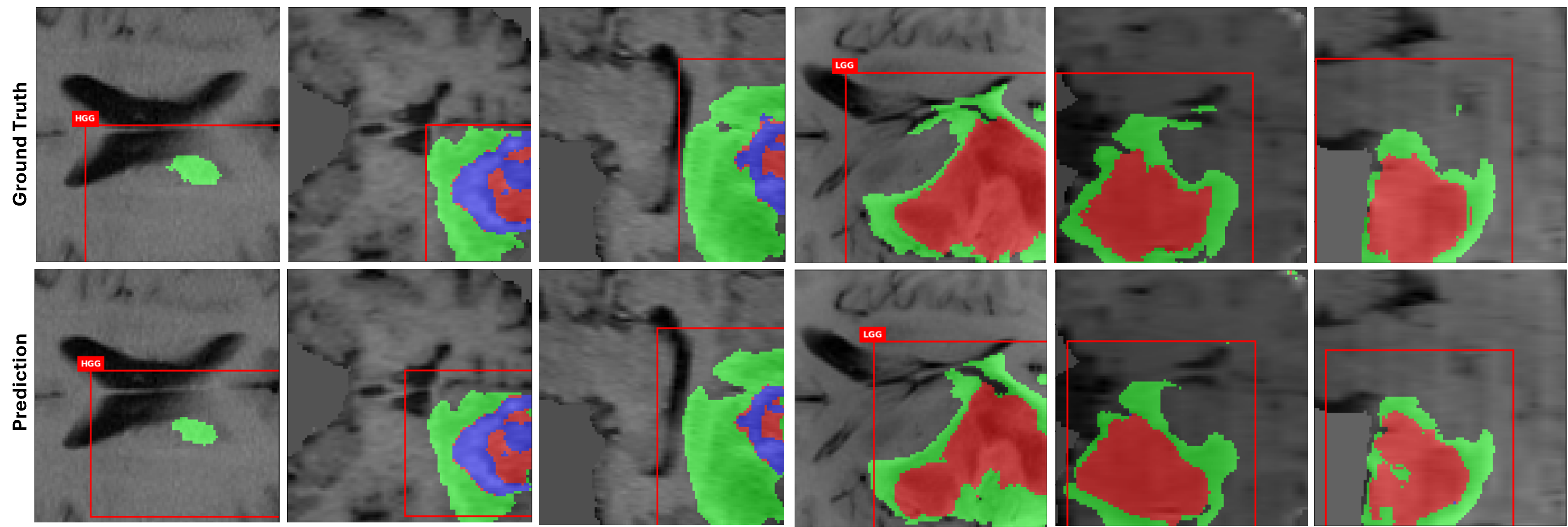}
\caption{Qualitative results for MTMed3D: The top row represents the Ground Truth, while the bottom row shows the outputs of MTMed3D. The left three columns depict HGG samples, representing three brain directions, and the right three columns show LGG samples. Red indicates the Tumor Core, green represents the Whole Tumor, and blue denotes the Enhanced Tumor. }
\label{fig_outputs}
\end{figure*}
\Cref{tab:fivefold_reults} reports the best and average results over five-fold cross-validation on BraTS 2018 and 2019 for detection, segmentation, and classification, where the best fold is defined as the one with the highest average performance across all three tasks. mAP@0.1:0.5 refers to the mAP computed over IoU thresholds from 0.1 to 0.5 with a step size of 0.05, while mAP@0.5 denotes the result at a fixed IoU of 0.5. 
Since the BraTS dataset was originally designed for segmentation and classification tasks, there are limited existing 3D detection methods evaluated on this dataset. To establish a fair comparison, we implemented a 3D RetinaNet with a ResNet-FPN backbone and trained it using the fold that achieved the best performance in our cross-validation experiments. This model served as our detection baseline. We also compared our proposed model against nnDetection and MedYOLO \cite{sobek2024medyolo}, though their reported results are available at mAP@0.5, a dash (“-”) indicates that no relevant data were available. \Cref{tab:det_performance} presents the detailed results, showing that MTMed3D outperforms these methods, possibly due to the hierarchical Transformer encoder's ability to effectively capture spatial context. The strong mAP and mAR values demonstrate the effectiveness of our model in accurately localizing brain tumors in 3D medical images. While our model outperforms existing methods on BraTS, further validation on additional datasets is necessary to assess its generalization capability in broader detection scenarios.
In \Cref{fig_outputs}, the red rectangular boxes represent the predicted detection bounding boxes.

\subsubsection{Segmentation Results}
\begin{table}[htbp]
  \centering
  \caption{Tumor segmentation performance of different methods.}
  \resizebox{\linewidth}{!}{
    \begin{tabular}{ccccccccc}
      \toprule
      {\textbf{Methods}} & \multicolumn{4}{c}{\textbf{Dice}} & \multicolumn{4}{c}{\textbf{HD}} \\
      \cmidrule(lr){2-5} \cmidrule(lr){6-9}
       & \textbf{WT} & \textbf{TC} & \textbf{ET} & \textbf{Avg.} & \textbf{WT} & \textbf{TC} & \textbf{ET} & \textbf{Avg.}\\
      \midrule
      AutoReg \cite{myronenko20193d} & 0.8839 & 0.8154 & 0.7664  & 0.8219 & 5.9044 & \textbf{4.8091} & 3.7731 & 4.8289\\
      MIC-DKFZ \cite{isensee2019no} & 0.8781 & 0.8062 & 0.7788 & 0.8210 & 6.0300 & 5.0800 & \textbf{2.9000} & \textbf{4.6700} \\
      DeepSCAN \cite{mckinley2019ensembles} & 0.8859 & 0.7992 & 0.7318 & 0.8056 & 5.5185 & 5.5347 & 3.4808 & 4.8447\\
      MultiDeep \cite{zhou2019learning} & 0.8842 & 0.7960 & 0.7775 & 0.8192 & 5.4681 & 6.8773 & 2.9366 & 5.0940\\
      \midrule
      MTMed3D (Best) & \textbf{0.9082} & \textbf{0.8183} & \textbf{0.8038} & \textbf{0.8434} & \textbf{5.4346} & 6.9773 & 5.6794  & 6.0304\\
      \bottomrule
    \end{tabular}
  }
  \label{tab:performance_metrics}
\end{table}
In \Cref{tab:fivefold_reults}, the model shows strong segmentation performance with high Dice and low HD across all tumor regions. \Cref{tab:performance_metrics} compares the segmentation performance of our model with several state-of-the-art (SOTA) methods leading in the BraTS 2018 challenge rankings. Since BraTS requires server submission for test evaluation and lacks detection labels, we report five-fold cross-validation results to ensure consistency across the three tasks. MTMed3D achieves the highest Dice scores. Although our model slightly lags behind some methods in HD for specific components, the overall performance of MTMed3D remains competitive.
\Cref{fig_outputs} presents the output results of our multi-task model. Comparing the ground truth with the segmentation results, while there are little differences in some areas, such as the boundary of the WT and TC, the overall results are closely aligned with the ground truth.

[\textbf{Insight 1}] The compared methods from the ranking are designed for single tasks and rely on ensembling to improve the performance. In contrast, MTMed3D achieves competitive results using a single end-to-end architecture, offering better efficiency and practicality for clinical deployment.

\subsubsection{Classification Results}
\Cref{tab:fivefold_reults} details MTMed3D's classification performance on the BraTS 2018 and 2019 datasets. Our average 5-fold cross-validation results show higher sensitivity but lower specificity, suggesting a bias toward the majority HGG class. We applied Focal Loss to address this class imbalance, but it did not lead to improved performance. \Cref{tab:classification_performance} compares the classification performance of our model with SOTA methods on the BraTS 2018 dataset. 
According to \Cref{tab:classification_performance}, other methods outperform ours. 
This is primarily because these methods operate on 2D slices rather than directly on 3D volumes and rely on dedicated preprocessing designed for classification.
For instance, Mask R-CNN\cite{zhuge2020automated}, CNN \cite{hafeez2023cnn}, and SE-ResNeXt \cite{linqi2022glioma} are all based on 2D inputs. However, our model is a multi-task model that processes all tasks on the same 3D data, without separate 2D preprocessing or classification steps. In addition, ConvNet \cite{zhuge2020automated} uses a two-stage pipeline that segments the tumor and then applies additional data augmentation to the segmented regions before training a classification model. MTMed3D is an end-to-end model, eliminating the need for manually designed intermediate steps and simplifying the overall training and inference pipeline. Moreover, SE-ResNeXt applies transfer learning to enhance generalization. In contrast, our multi-task model not only ensures the performance of each task but also maintains a balance across tasks, preventing any single task from dominating training. This constraint limits the use of task-specific optimization or pre-trained weights that could disproportionately favor one task over others.
\Cref{fig_outputs} shows classification results above the red bounding boxes, with both HGG and LGG correctly classified.
\begin{table}[htbp]
  \centering
  \caption{Classification performance of different methods.}
  \resizebox{\linewidth}{!}{
      \begin{tabular}{lccc}
        \toprule
        \textbf{Methods} & \textbf{Accuracy} & \textbf{Sensitivity} & \textbf{Specificity} \\
        \midrule
        Mask R-CNN (2D) \cite{zhuge2020automated} & 0.9630 & 0.9350 & 0.9720 \\
        ConvNet (3D) \cite{zhuge2020automated} & 0.9710 & 0.9470 & 0.9680 \\
        CNN (2D) \cite{hafeez2023cnn} & 0.9715 & 0.9818 & \textbf{0.9855} \\
        SE-ResNeXt (2D) \cite{linqi2022glioma} & \textbf{0.9745} & \textbf{0.9835} & 0.9493 \\
        \midrule
        MTMed3D (Best) & 0.9298 & 0.9286 & 0.9333 \\
        \bottomrule
      \end{tabular}
    }
      \label{tab:classification_performance}
\end{table}

[\textbf{Insight 2}] 
\begin{table*}[htbp]
  \centering
  \caption{Performance Comparison of Multi-Task and Single-Task Models and Ablation Study on Multi-Task Components.}
  \resizebox{\textwidth}{!}{
    \begin{tabular}{cccccccccccccccc}
      \toprule
      \textbf{Task} 
      & \multicolumn{8}{c}{\textbf{Segmentation}} 
      & \multicolumn{3}{c}{\textbf{Classification}} 
      & \multicolumn{2}{c}{\textbf{Detection}} 
      & \textbf{Task Avg.} \\
      \cmidrule(r){2-9} \cmidrule(r){10-12} \cmidrule(r){13-14}
      & \multicolumn{4}{c}{\textbf{Dice}} 
      & \multicolumn{4}{c}{\textbf{HD}} 
      & \textbf{Acc} & \textbf{Sen} & \textbf{Spe} 
      & \textbf{mAP@} & \textbf{mAR@} & \\
      \cmidrule(r){2-5} \cmidrule(r){6-9}
      \textbf{Region} 
      & \textbf{WT} & \textbf{TC} & \textbf{ET} & \textbf{Avg.} 
      & \textbf{WT} & \textbf{TC} & \textbf{ET} & \textbf{Avg.} 
      &  &  &  
      & 0.1:0.5 & 0.1:0.5 & \\
      \midrule
      MTMed3D (Avg.) 
      & 0.8793 & 0.8019 & 0.7755 & 0.8189 
      & 9.7864 & 8.3361 & 6.7452 & 8.2892 
      & 0.9193 & 0.9761 & 0.7634 
      & 0.9082 & 0.9357 & \textbf{0.8955} \\
      Single-task (Avg.) 
      & 0.8837 & 0.7972 & 0.7348 & 0.8052 
      & 6.7715 & 6.2509 & 5.4199 & 6.1474 
      & 0.9017 & 0.9594 & 0.7687 
      & 0.8217 & 0.8608 & 0.8474 \\
      \midrule
      FPN + GradNorm (Avg.) 
      & 0.8803 & 0.8060 & 0.7582 & 0.8148 
      & 10.2574 & 8.7848 & 6.9169 & 8.6530 
      & 0.9123 & 0.9859 & 0.7056 
      & 0.8963 & 0.9310 & 0.8886 \\
      FPN + MGDA (Avg.) 
      & 0.8826 & 0.8070 & 0.7837 & 0.8244 
      & 9.1462 & 8.0395 & 6.4389 & 7.8749 
      & 0.8666 & 0.9712 & 0.5760 
      & 0.9172 & 0.9399 & 0.8870 \\
      PANet + MGDA (Avg.) 
      & 0.8772 & 0.8036 & 0.7763 & 0.8190 
      & 10.1070 & 8.2520 & 6.6129 & 8.3240 
      & 0.8632 & 0.9812 & 0.5401 
      & 0.9070 & 0.9380 & 0.8818 \\
      \bottomrule
    \end{tabular}
  }
  \label{tab:multivssingle}
\end{table*}
\begin{table}[htbp]
  \centering
  \caption{Efficiency comparison between MTMed3D and three single-task models ($\downarrow$ indicates lower is better).}
  \resizebox{\linewidth}{!}{
      \begin{tabular}{lccccc}
        \toprule
        \textbf{Model Name} & \textbf{MACs $\downarrow$} & \textbf{FLOPs $\downarrow$} & \textbf{Params $\downarrow$} & \textbf{Inference (s) $\downarrow$} & \textbf{Size (MB) $\downarrow$} \\
        \midrule
        Single-task Models & $4 \times 10^{11}$ & $7 \times 10^{11}$ & $1 \times 10^{8}$ & 0.2248 & 587.81 \\
         MTMed3D & \bm{$2 \times 10^{11}$} & \bm{$4 \times 10^{11}$} & \bm{$8 \times 10^{7}$} & \textbf{0.0996} & \textbf{306.79} \\
        \bottomrule
      \end{tabular}
    }
  \label{tab:efficiency_comparison}
\end{table} 
While MTMed3D performs well on segmentation and detection tasks, classification remains more challenging, possibly due to the absence of a learnable class token, which may limit its ability to represent features critical for distinguishing LGG from HGG. Although our classification performance does not surpass other methods, it suggests Swin Transformer may be insufficient in glioma grading classification without additional structural enhancements, which is also reflected in the ResMT \cite{cui2024resmt}, a model trained on BraTS 2019 that incorporates a parallel CNN branch alongside Swin UNETR and is further supplemented by additional modules to improve classification.

\subsubsection{Multi-task vs Single-task Models}
\Cref{tab:multivssingle} compares the performance of MTMed3D and single-task models. The Task Avg. column represents the average of Dice, accuracy, and detection performance, providing an overall measure across all tasks. Each single-task model follows the same architecture as its multi-task counterpart but is trained independently. While the single-task model achieves better segmentation results in terms of HD, MTMed3D remains competitive. In classification, MTMed3D demonstrates higher accuracy and sensitivity, while for detection, it has higher mAP and mAR. 
We also compared the efficiency of our MTMed3D with three single-task models, as illustrated in \Cref{tab:efficiency_comparison}. MTMed3D significantly reduces computational costs, with approximately 46.64\% fewer MACs and FLOPs. It also requires 47.80\% fewer parameters and less size. The inference time is reduced by 55.70\%, enabling faster decision-making. This significant reduction in computational demand, along with its strong performance, makes MTMed3D a practical choice for real-world clinical settings where efficiency and accuracy are crucial under limited computing resources. Further analysis of the computation breakdown and inference time of different components is provided in Supplementary Section 3. Additionally, MTMed3D’s flexible design lets users activate tasks independently based on resources and clinical needs, optimizing efficiency without compromising accuracy. It directly processes 3D images, eliminating the need to convert 3D data into 2D slices and thus avoiding the overhead of repetitive forward propagation for each slice and the computational burden of reassembling the final output.

[\textbf{Insight 3}] 
 Comparing the multi-task and single-task models, we find that multi-task learning enhances detection performance by leveraging shared representations across tasks. Features crucial for detection, which may be difficult to learn in a single-task model, are more easily learned in a multi-task setting, as they also contribute to segmentation and classification. This shared learning process improves overall model performance.

\subsection{Ablation Studies}
In our ablation studies, we analyze the impact of different decoder designs and the task-balancing optimizer on performance. Additional analyses, including the effect of varying encoder depth and task interactions when training without the optimizer, are provided in Supplementary Sections 1 and 2.
\subsubsection{Detection Branch}
To evaluate the efficiency of PANet, we conducted an ablation study on different detection branches. \Cref{tab:multivssingle} shows results obtained using FPN as the detection branch. Although FPN (FPN + GradNorm) achieves a slightly higher Dice score for the WT and TC regions, MTMed3D (PANet + GradNorm) provides better performance in Dice for ET and consistently lower HD values. MTMed3D shows improved classification accuracy, specificity, and detection performance. Overall, PANet outperforms FPN.

\subsubsection{Optimizers}
In the optimizer, we utilize the GradNorm method to adjust gradient magnitudes by tuning multi-task loss functions. To evaluate the optimization strategy, we design an ablation study comparing GradNorm with the Multiple Gradient Descent Algorithm (MGDA) proposed by \cite{sener2018multi}. MGDA formalizes multi-task learning as a multi-objective optimization problem, aiming for Pareto optimality where no task's loss can be improved without harming others. Our comparative analysis reveals how these distinct approaches impact task performance trade-offs. \Cref{tab:multivssingle} compares MTMed3D with PANet + MGDA, showing that MGDA slightly outperforms in segmentation metrics but suffers from lower classification accuracy, highlighting an imbalance across tasks. In contrast, MTMed3D excels in classification accuracy and specificity, achieving the best balance and ensuring competitive results in segmentation and detection without compromising classification. Similarly, FPN + MGDA performs well in segmentation but struggles with classification. Overall, MTMed3D handles all tasks effectively without sacrificing performance.

\section{CONCLUSION}
We present MTMed3D, a novel end-to-end multi-task model for 3D detection, segmentation, and classification. Our multi-task design improves resource efficiency compared to single-task models that train tasks separately.
MTMed3D shows promising results on all three tasks, demonstrating not only its effectiveness but also the potential of the Transformer framework for multi-task learning in medical imaging applications. In particular, our detection results not only outperform existing methods on BraTS, but also show improvement over the single-task model under joint training, suggesting that multi-task learning with Transformer backbones offers benefits for detection.
While we primarily validated it on brain tumor datasets, it is adaptable to other datasets, making it suitable for various medical imaging scenarios and tasks. Overall, MTMed3D presents an efficient solution for multi-task medical image analysis, with the potential to serve as an assistive tool for radiologists to help alleviate workload.

\section*{References}
\vspace{-1.8em}
\bibliographystyle{IEEEtran}
\bibliography{references} 

\end{document}